\documentclass[pra,letterpaper,aps,floatfix,superscriptaddress,twocolumn]{revtex4}
\usepackage{graphicx}
\usepackage{amsmath,amssymb}
\usepackage{bm}

\def\ket#1{|#1\rangle }

\usepackage[pdfstartview=FitH,breaklinks=true,bookmarks=true,colorlinks=true,anchorcolor=black,citecolor=red,filecolor=black,menucolor=black,urlcolor=blue,linkcolor=blue]{hyperref}

\begin{document}

\title{Fate of Fermi-arc States in Gapped Weyl Semimetals under Long-ranged Interactions}

\author{Yu-Chin Tzeng}
\affiliation{Department of Physics, National Chung Hsing University, Taichung 40227, Taiwan}
\author{Min-Fong Yang}
\email{mfyang@thu.edu.tw}
\affiliation{Department of Applied Physics, Tunghai University, Taichung 40704, Taiwan}

\date{\today}

\begin{abstract}
For noninteracting Weyl semimetals (WSMs), Fermi-arc surface states can arise when there exist gapless Weyl nodes in the bulk single-particle spectrum. However, in the presence of electronic correlations, it is not clear whether this bulk-boundary correspondence still holds or not. Recently, novel correlated phases are predicted to appear in WSMs with long-ranged interactions, in which the bulk Weyl nodes can be gapped but without destroying their topological properties. Here, we explore the fate of the Fermi-arc states under the influence of such long-ranged interactions.
After mapping the system onto a one-dimensional interacting Su-Schrieffer-Heeger (SSH) model with two open ends, we employ numerical exact diagonalizations to address the issue whether the Fermi-arc states will be modified. By extrapolating our data to the thermodynamic limit, we find that the zero-energy edge states of the corresponding SSH model still exist for those momenta giving the noninteracting Fermi-arc states. Moreover, this observation applies to both the single-particle and the collective edge excitations. Since the locus of these edge states constitutes the Fermi arcs, the robustness of the Fermi-arc states against long-ranged interactions is thus demonstrated. In particular, the Fermi arcs of single-particle nature can survive even when the single-particle gaps at the Weyl nodes are opened by interactions.
Our results illustrate the subtlety in identifying the topological phases of interacting WSMs and show the limitation of the approaches simply by examining the nodal structure of the single-particle spectrum.
\end{abstract}

\maketitle

\section{introduction}

There is ongoing interest in understanding the effects of electronic correlations on topological quantum states owing to the promise of uncovering new phenomena in condensed matter~\cite{Budich-Trauzettel2013,Hohenadler-Assaad2013,Meng_etal2014,Rachel18}. Among such topological phases of matter, Weyl semimetals (WSMs) represent an nontrivial class as they broaden the topological classification of matter beyond the insulators~\cite{Hasan_etal2017,Yan-Felser2017,Burkov2018,Armitage_etal2018}.
In the absence of interaction, the bulk system of WSMs hosts gapless band-touching points, called Weyl nodes. Each of these nodal points acts as a monopole of the Berry curvature in the momentum space with a definite Chern number, which is proportional to the quantized Berry flux around the node. Another key signature of WSMs is the presence of Fermi arcs that connect pairs of projected Weyl nodes with opposite Chern numbers in the surface Brillouin zone~\cite{Wan_etal2011}. Actually, as a consequence of the bulk-boundary correspondence, the existence of the surface Fermi arcs is guaranteed by the nontrivial topology of the Berry curvature carried by the bulk Weyl nodes.
The fascinating physics in WSMs has been understood well at the level of noninteracting description. Thereafter, one of the subjects that need to be further explored is the correlation effects on WSMs.

The robustness of the above features of WSMs against electronic correlations is being actively pursued~\cite{Witczak-Krempa_etal2014,Roy_etal2017,Carlstrom-Bergholtz2018-1,%
Carlstrom-Bergholtz2018-2,Acheche_etal2019,Crippa_etal2020,%
Laubach_etal2016,Wang_etal19,Morimoto-Nagaosa2016,Meng-Budich2019,%
Zhang_Zubkov19,Xu_etal19}.
Based on perturbative approaches, weak short-range interactions was shown to provide merely band renormalization and WSMs are thus stable~\cite{Witczak-Krempa_etal2014,Roy_etal2017,Carlstrom-Bergholtz2018-1,%
Carlstrom-Bergholtz2018-2}. Nevertheless, transitions to insulating states can be induced by sufficiently strong interactions~\cite{Roy_etal2017,Acheche_etal2019,Crippa_etal2020}.
Within the nonperturbative mean-field theories, it was found that, in the presence of either the spin density wave~\cite{Laubach_etal2016} or the Fulde-Ferrell-Larkin-Ovchinnikov superconducting~\cite{Wang_etal19} instability, the bulk Weyl nodes can be gapped out, while the Fermi arcs keep unchanged. This is achieved by pair annihilation of Weyl nodes coming from the backfolding of the Brillouin zone caused by ordering. Therefore, contrary to the noninteracting case, the emergence of the surface Fermi arcs needs not be associated with the existence of the gapless Weyl nodes.

Nevertheless, the validity of usual approximations for exploring the strongly correlated phases is not so clear. Thus it is worthwhile to study exactly solvable models for searching possible novel physics in the strongly correlated regime. Recently, exactly solvable models of interacting WSMs have been investigated in Refs.~\cite{Morimoto-Nagaosa2016,Meng-Budich2019}, where the  two-body interactions are chosen to be local in momentum space and then of infinite range in real space. The advantage of such models is that different momentum states are decoupled and they thus can be solved exactly through diagonalization for each momentum state.
Distinct from the mechanism discussed in Refs.~\cite{Laubach_etal2016,Wang_etal19}, it was shown that the Weyl nodes in these models can be gapped out by interactions but without the pair annihilation. Notably, the topological properties of noninteracting WSMs are kept unchanged: these gapped Weyl nodes still carry nonzero Chern numbers with their noninteracting values,  and the Fermi arcs on the surfaces persist against the correlation effects~\cite{Morimoto-Nagaosa2016}. This novel phase behaves as a topological Mott insulator and is dubbed as the Weyl Mott insulator.

We note that, contrary to their bulk behaviors, the surface properties (say, the Fermi arcs) of these solvable models can not be derived analytically. It is because the translational symmetry is broken in the presence of boundaries and the momenta are not good numbers any more.
Usually, the Fermi arc states of the interacting WSMs are calculated by considering an equivalent single-particle Hamiltonian $H_t$ (i.e., the so-called ``topological Hamiltonian"~\cite{Wang-Zhang2012,Wang-Yan2013}) under open boundary conditions~\cite{Morimoto-Nagaosa2016,Xu_etal19}. However, as pointed out in the context of topological insulators, the eigenstates of $H_t$ have no direct physical meaning~\cite{Gurarie2011,Essin-Gurarie2011}. Therefore, the Fermi-arc states derived from $H_t$ may not be real.
In order to provide soild supports on the existence of the Fermi arcs in interacting WSMs, approaches other than that using $H_t$ are thus necessary.

Moreover, as mentioned in Ref.~\cite{Yang2019}, the surface Fermi-arc states directly implied by quantized Hall conductances (i.e., quantized Chern numbers) consist of \emph{bosonic charge-neutral} particle-hole excitations, rather than those of single-particle nature. This conclusion is drawn from the theory of quantum Hall edge states~\cite{Wen1991,Renn1995,Yoshioka}, since each two-dimensional (2D) momentum sector between a pair of Weyl nodes can be viewed as a quantum anomalous Hall insulator. That is, the topologically induced Fermi-arc states are the collective edge excitations, whose appearance cannot in general be implied by solving the effective single-particle Hamiltonian $H_t$.

In this paper, we revisit the issue of the correlation effects on surface Fermi-arc states of WSMs with long-ranged two-body interactions introduced in Refs.~\cite{Morimoto-Nagaosa2016,Meng-Budich2019}. In the noninteracting setting, a single pair of gapless Weyl nodes separated by a distance along the $k_z$ direction in the momentum space is assumed. In order to construct the Fermi-arc states, we consider a finite-slab geometry with open boundary conditions in the $x$ direction, but keeping translational symmetry along the other two transverse directions. The transverse momenta
$\mathbf{k}_\bot\equiv(k_y,\,k_z)$ thus remain good quantum numbers. Utilizing the fact that different momentum states are decoupled in the present case, we can solve our system once for a given $\mathbf{k}_\bot$. As a consequence, our problem reduces to a one-dimensional (1D) interacting Su-Schrieffer-Heeger model~\cite{Su_etal1979} in the $x$ direction with two open ends.

By means of exact diagonalization, we search for the edge modes within both the single-particle and the particle-hole subspaces of the Hilbert space. The single-particle edge modes are identified by calculating the energy and the density distribution of the quasi-particle added or removed from the corresponding 1D system at half-filling. Interestingly, we find that the zero-energy edge modes of single-particle excitations still appear in the spectrum, even though the long-ranged interactions induce single-particle gaps at the nodes. Besides, these edge modes occur at the same positions on the surface Brillouin zone as those for the noninteracting Fermi arcs. Actually, there exist tiny single-particle excitation gaps for finite systems. However, their values reduce to zero after extrapolating to the thermodynamic limit. Because the single-particle Fermi arcs are constituted by the collection of such zero-energy edge states, our findings provide direct evidences on their robustness against long-ranged interactions.

On the other hand, the collective edge modes can be recognized by evaluating the energy and the density distribution of the particle-hole excitations with $\Delta\mathbf{k}_\bot=0$. As mentioned above, the zero-energy edge modes among these particle-hole excitations represent the quantum Hall edge states implied by the quantized Hall conductances. Our data show that, the same as those single-particle surface states, the collective Fermi-arc states of particle-hole nature persist to exist as well even in the presence of long-ranged interactions. The extent of these zero-energy collective modes on the surface Brillouin zone is in agreement with the calculated results of the bulk Hall conductances~\cite{Morimoto-Nagaosa2016}. That is, collective edge states will show up for those $k_z$'s such that the bulk Hall conductances (i.e., Chern numbers) keep quantized. Our findings thus indicate that the stability of the collective Fermi-arc surface states originates from the nontrivial topology encoded in the particle-hole subspace, no matter whether the single-particle gaps at the nodes are opened or not.

This paper is organized as follows. Our model Hamiltonian is introduced in Sec.~II and some of its bulk properties are summarized. In the second part of this section, the 1D interacting Su-Schrieffer-Heeger model for the study of surface Fermi arcs is derived. Our numerical results are presented in Sec.~III. They provide clear evidences for the robustness of the Fermi arcs against long-ranged interactions. We summarize our work in Sec.~IV.

\section{model Hamiltonian}

For simplicity, we consider a typical two-band lattice model of WSMs [see Eq.~\eqref{eq:H}], which host a single pair of Weyl nodes located at momenta $\mathbf{k}=(0,\,0,\,\pm k_\mathrm{Weyl})$ in the noninteracting setting. For a fixed $k_z$ with $|k_z|<k_\mathrm{Weyl}$, the 2D slice of the Brillouin zone carries a nonzero Chern number $C=1$ and shows the quantized anomalous Hall effect, while $C=0$ otherwise. The region between nodes can thus be viewed as a stack of 2D Chern insulators. If we consider a finite geometry along the direction perpendicular to the $z$ axis, there appear chiral edge states near the Fermi energy on the boundaries of these 2D Chern insulators. The collection or locus of such edge states constitutes the Fermi arcs, which terminate at the projection of the bulk Weyl nodes onto the surface Brillouin zone~\cite{Wan_etal2011}.

Following Refs.~\cite{Morimoto-Nagaosa2016,Meng-Budich2019}, we explore the influence of the momentum-local two-body interactions on the surface Fermi arcs of such a two-band model of WSMs. Let's begin with the description of our system and a summary on some of its bulk properties.

\subsection{Bulk excitation energies}\label{bulk}

Our model Hamiltonian under periodic boundary conditions is given by
\begin{eqnarray}
H &=& H_0 + H_U \; , \label{eq:H}\\
H_0 &=& \sum_{\mathbf{k}} \sum_{\alpha,\beta}
c^\dag_{\mathbf{k}\alpha} [\mathbf{h}(\mathbf{k})\cdot\bm{\sigma}]_{\alpha\beta} c_{\mathbf{k}\beta} \; , \label{eq:H0}\\
H_U &=& \frac{U}{2} \sum_{\mathbf{k}}
\left( n_{\mathbf{k}\uparrow} + n_{\mathbf{k}\downarrow} - 1 \right)^2
\nonumber \\
&=& U \sum_\mathbf{k} n_{\mathbf{k}\uparrow}n_{\mathbf{k}\downarrow}
-\frac{U}{2} \sum_\mathbf{k} (n_{\mathbf{k}\uparrow}+n_{\mathbf{k}\downarrow}-1) \; . \label{eq:HU}
\end{eqnarray}
Here $\bm{\sigma}$ is the vector of three Pauli matrices denoting a (pseudo-)spin-$1/2$ operator, $c^\dag_{\mathbf{k}\alpha}$ represents the creation operator for electrons of (pseudo-)spin $\alpha=\;\uparrow$ or $\downarrow$, and $n_{\mathbf{k}\alpha}=c^\dag_{\mathbf{k}\alpha}c_{\mathbf{k}\alpha}$. The first part $H_0$ describes a noninteracting two-band model of a WSM. To be specific, we take $\mathbf{h}(\mathbf{k})=(M-\cos k_x -\cos k_y -\cos k_z,\;\sin k_x,\;\sin k_y)$ with $1<M<3$~\cite{note1}. In such a case, only a single pair of Weyl nodes appears at momenta $\mathbf{k}=(0,\,0,\,\pm k_\mathrm{Weyl})$ with $k_\mathrm{Weyl}=\arccos(M-2)$.
The second term $H_U$ gives a Hubbard-like interaction but being local in momentum space, rather than in real space.

Because different momentum states are decoupled, the bulk properties of this interacting model can be derived exactly through diagonalizing the Hamiltonian in the number representation for each momentum $\mathbf{k}$. The four eigenstates for each $\mathbf{k}$ are $\{ \ket{0},\; b_{\mathbf{k}-}^\dag\ket{0},\; b_{\mathbf{k}+}^\dag\ket{0},\; b_{\mathbf{k}-}^\dag b_{\mathbf{k}+}^\dag\ket{0} \}$ with the corresponding eigenvalues $\{ U/2,\; -h(\mathbf{k}),\; h(\mathbf{k}),\; U/2 \}$. Here $b^\dag_{\mathbf{k}+}$ and $b^\dag_{\mathbf{k}-}$ denote the eigenmodes of the noninteracting Hamiltonian $H_0$, $\ket{0}$ is the vacuum state, and $h(\mathbf{k}) = | \mathbf{h}(\mathbf{k})|$. Therefore, the many-body ground state for positive $U$ at zero temperature and half-filling becomes the state with a completely-filled lower band, $|\Psi_0\rangle=\prod_\mathbf{k} b_{\mathbf{k}-}^\dag\ket{0}$.
The single-particle/hole excitations, $b_{\mathbf{k}+}^\dag \ket{\Psi_0}$ and $b_{\mathbf{k}-}\ket{\Psi_0}$, have nonzero excitation energies $\Delta E_{1p/1h}=h(\mathbf{k})+U/2$ even at the Weyl nodes such that $h(\mathbf{k})=0$. The system thus behaves as a Mott insulator with a full gap.
However, we note that the excitation energies of the \emph{bosonic charge-neutral} particle-hole excitations, $b_{\mathbf{k}+}^\dag b_{\mathbf{k}-}\ket{\Psi_0}$, keep their noninteracting values $\Delta E_\mathrm{ph}=2h(\mathbf{k})$, which become zero at the Weyl nodes. That is, while the single-particle/hole excitations are always gapful, this interacting WSM retains its \emph{gapless} Weyl nodes from the viewpoint of particle-hole excitations.

\subsection{One-dimensional model for the study of surface Fermi arcs}

In order to construct the Fermi arc surface states, we need to consider a finite-slab geometry with open boundaries. To this end, we first transform Eq.~\eqref{eq:H} into its real-space form and then impose the open boundary conditions. Here we take a finite size of $L$ sites in the $x$ direction. By employing Fourier transform in the $x$ direction, $\tilde{c}_{j\mathbf{k}_\bot\alpha}=(1/\sqrt{L})\sum_{k_x} e^{ik_x j} c_{\mathbf{k}\alpha}\,$, the Hamiltonian for a given transverse momentum $\mathbf{k}_\bot\equiv(k_y,\,k_z)$ reduces to a 1D tight-binding form, where $j=1,\dots,L$ labels the sites in the $x$ direction. After imposing the the open boundary conditions by dropping off the boundary hopping terms, the resulting 1D model with two open ends becomes
\begin{equation}\label{eq:1D_H}
\mathcal{H}(\mathbf{k}_\bot) =\mathcal{H}_0 (\mathbf{k}_\bot)
+ \mathcal{H}_U (\mathbf{k}_\bot)
\end{equation}
with
\begin{align}\label{eq:1D_H0}
\mathcal{H}_0 (\mathbf{k}_\bot)= &-\sum_{i=1}^{L-1} \left(\tilde{c}_{i\uparrow}^\dagger \tilde{c}_{i+1\downarrow} + \mathrm{h.c.}\right)
+t_{\mathbf{k}_\bot}\sum_{i=1}^L \left(\tilde{c}_{i\uparrow}^\dagger \tilde{c}_{i\downarrow} + \mathrm{h.c.}\right) \nonumber \\
&+h_{\mathbf{k}_\bot}\sum_{i=1}^L \left(\tilde{c}_{i\uparrow}^\dagger \tilde{c}_{i\uparrow} - \tilde{c}_{i\downarrow}^\dagger \tilde{c}_{i\downarrow}\right)
\end{align}
and
\begin{align}\label{eq:1D_HU}
\mathcal{H}_U (\mathbf{k}_\bot) =
&\frac{U}{L}\sum_{d=1}^{L-1}\sum_{i=1}^{L-d}\sum_{j=d+1}^L
\left[ \tilde{c}_{i+d\uparrow}^\dagger \tilde{c}_{i\uparrow} \tilde{c}_{j-d\downarrow}^\dagger \tilde{c}_{j\downarrow} +(\uparrow\leftrightarrow\downarrow) \right] \nonumber \\
&+\frac{U}{L}\sum_{i,j=1}^L \tilde{n}_{i\uparrow} \tilde{n}_{j\downarrow}
-\frac{U}{2}\sum_{i=1}^L ( \tilde{n}_{i\uparrow} + \tilde{n}_{i\downarrow} -1 ) \; .
\end{align}
Here the labels of the transverse momentum $\mathbf{k}_\bot$ in the fermion operators is suppressed for simplification and $\tilde{n}_{i\alpha}=\tilde{c}_{i\alpha}^\dagger \tilde{c}_{i\alpha}$. The two parameters in Eq.~\eqref{eq:1D_H0} are given by $t_{\mathbf{k}_\bot}=M-\cos k_y -\cos k_z$ and $h_{\mathbf{k}_\bot}=\sin k_y$.
We note that the non-interacting part $\mathcal{H}_0(\mathbf{k}_\bot)$ is equivalent to the Su-Schrieffer-Heeger model~\cite{Su_etal1979} with an alternating on-site potential.
The first term of $\mathcal{H}_U(\mathbf{k}_\bot)$ describes the correlated hopping for up and down (pseudo-)spins with the same distances $d$ but in opposite directions. The second term in Eq.~\eqref{eq:1D_HU} comes from the zero-distance $d=0$ hopping, which reduces to the density-density interactions between different spin components.

While the model in Eq.~\eqref{eq:H} under periodic boundary conditions can be solved analytically, the corresponding one under open boundary conditions cannot. In this paper, numerically exact diagonalization is employed to search possible edge states at zero energy for the corresponding 1D models in Eq.~\eqref{eq:1D_H}. The collection of such zero-energy edge states constitutes the surface Fermi arcs of our interacting WSM.

\section{results}

In the present calculations, we set $M=2$ such that $k_\mathrm{Weyl}=\pi/2$. The system sizes under consideration are up to $L=16$. In the following, we focus on the $U=2$ case. The results for other $U$'s are qualitatively the same.

Due to the conservation of total particle number, subspaces of the Hilbert space with different particle numbers are decoupled. One can thus diagonalize $\mathcal{H}(\mathbf{k}_\bot)$ for each subspace with a fixed number $N$ of electrons. The largest subspace takes place at half-filling $N=L$ and its dimension is about $6\times10^8$ for $L=16$.

\subsection{Single-particle Fermi arcs}\label{1p_arc}

To show the existence of the single-particle edge states lying between two bulk bands, the quasi-particle and the quasi-hole energies, defined by $\epsilon_+\equiv \mathcal{E}_0(N=L+1)-\mathcal{E}_0(N=L)$ and $\epsilon_-\equiv \mathcal{E}_0(N=L)-\mathcal{E}_0(N=L-1)$, are calculated. Here $\mathcal{E}_0(N)$ denotes the ground energy of the 1D model in Eq.~\eqref{eq:1D_H} with $N$ electrons. The appearance of the in-gap edge states can be recognized when $\epsilon_+$ and $\epsilon_-$ deviate from their bulk values,
\begin{align}
\epsilon^\mathrm{bulk}_\pm &=\pm\min_{k_x}\{\Delta E_{1p/1h}\} \nonumber \\ &=\pm\left[\sqrt{\sin^2{k_y}+(1-\cos{k_y}-\cos{k_z})^2}+U/2\right] \; , \nonumber
\end{align}
for the present case of $M=2$.

Before presenting our numerical results, some symmetry properties of our 1D lattice model $\mathcal{H}(\mathbf{k}_\bot)$ in Eq.~\eqref{eq:1D_H} are discussed in order.
First, $\mathcal{H}(\mathbf{k}_\bot)$ is invariant under the parity transformation $\mathcal{P}$: $\tilde{c}_{i\alpha}\rightarrow\tilde{c}_{L-i+1,-\alpha}$ in combination with $k_y\rightarrow -k_y$. The energy spectrum will thus be symmetric with respect to the $k_y=0$ axis.
Second, $\mathcal{H}(\mathbf{k}_\bot)$ is invariant under the operation combining the parity transformation $\mathcal{P}$ with the particle-hole transformation: $\tilde{c}^\dagger_{i\uparrow}\leftrightarrow\tilde{c}_{i\uparrow}$ and $\tilde{c}^\dagger_{i\downarrow}\leftrightarrow-\tilde{c}_{i\downarrow}$. Therefore, the quasi-particle and the quasi-hole excitation energies, $\epsilon_+$ and $|\epsilon_-|$, above the ground-state energy will be the same. That is, the energy spectrum will become symmetric with respect to $\epsilon=0$.

\begin{figure}[tp]
\includegraphics[width=0.48\textwidth]{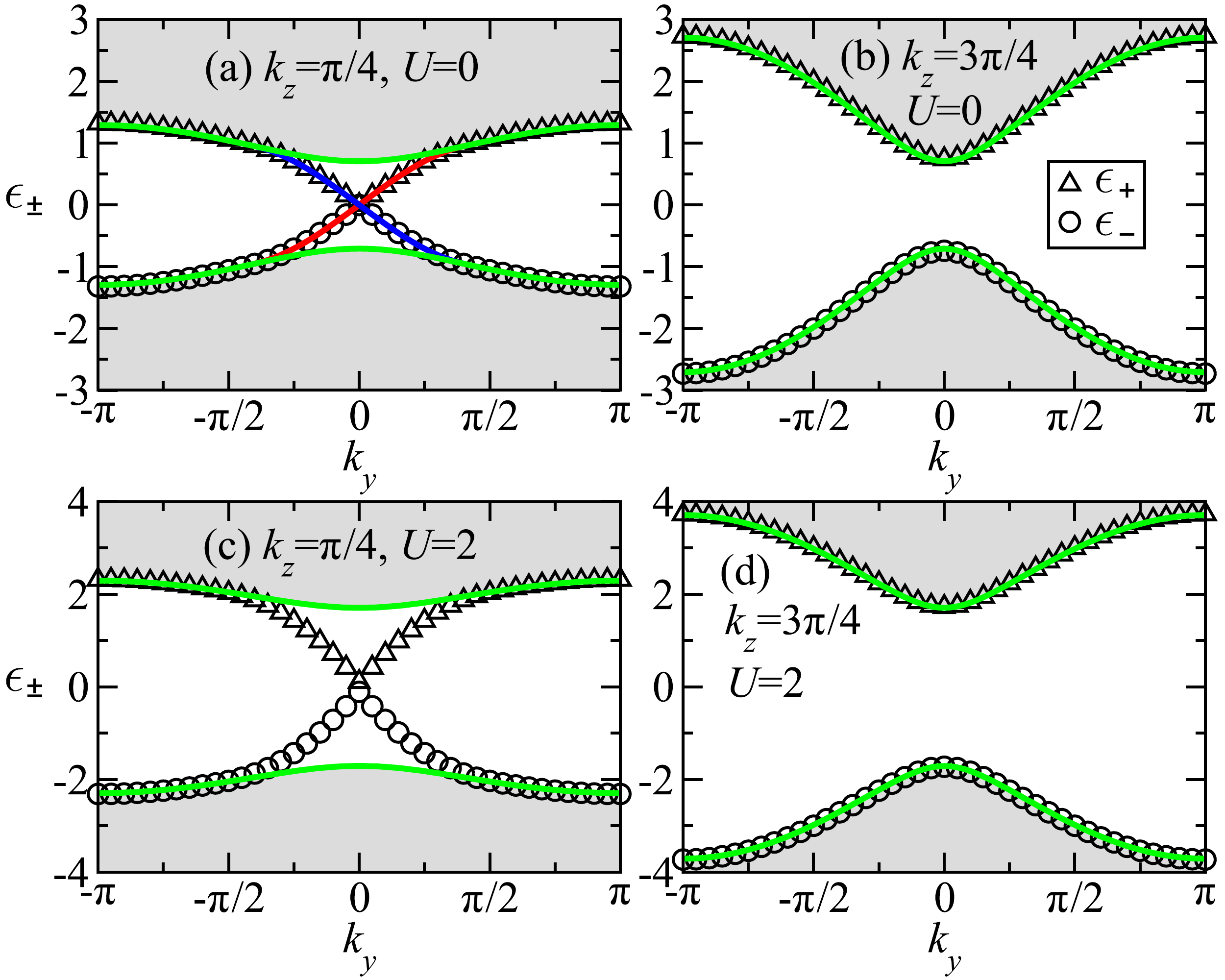}
\caption{Quasi-particle and quasi-hole energies, $\epsilon_+$ and $\epsilon_-$, at two different $U$'s as functions of $k_y$ for $k_z=\pi/4$($<k_\mathrm{Weyl}$) and $3\pi/4$($>k_\mathrm{Weyl}$). Here the system size $L=16$ and $k_\mathrm{Weyl}=\pi/2$ for $M=2$. The upper panels show the $U=0$ cases, while the lower ones give those for $U=2$.
Black open triangles represent $\epsilon_+$ and black open circles show $\epsilon_-$. The green lines denote the corresponding bulk values $\epsilon^\mathrm{bulk}_\pm$. The shadow regions indicate the quasi-particle/hole bands bounded by $\epsilon^\mathrm{bulk}_\pm$. For $U=0$ case with $k_z=\pi/4$, the red (blue) line shows the dispersion relation of the noninteracting edge states, $\epsilon=\sin{k_y}$ ($\epsilon=-\sin{k_y}$), for $|k_y|\le0.4\pi$.}\label{1part_eng}
\end{figure}

In Fig.~\ref{1part_eng}, $\epsilon_+$ and $\epsilon_-$ as functions of $k_y$ for $k_z=\pi/4$ and $3\pi/4$ are displayed for the case of $U=2$. For comparison, the results for the noninteracting ($U=0$) case are shown as well. It is clear that our findings do respect the symmetries mentioned above. The $U=0$ results can be derived analytically. When $|k_z|<k_\mathrm{Weyl}$, the effective 2D system at a fixed $k_z$ acts as a Chern insulator with a Chern number $C=1$ and thus supports a pair of chiral edge modes. The edge modes emerge around $k_y=0$ with the dispersions $\epsilon=\pm\sin{k_y}$. On the other hand, when $k_\mathrm{Weyl}<|k_z|<\pi$, the effective 2D system behaves as a $C=0$ trivial insulator, in which there exists no edge state and the quasi-particle/quasi-hole energies must equal to their bulk values. As seen from Fig.~\ref{1part_eng}, our numerical data for the $U=0$ case fairly reproduce the analytic predictions.

\begin{figure}[tp]
\includegraphics[width=0.48\textwidth]{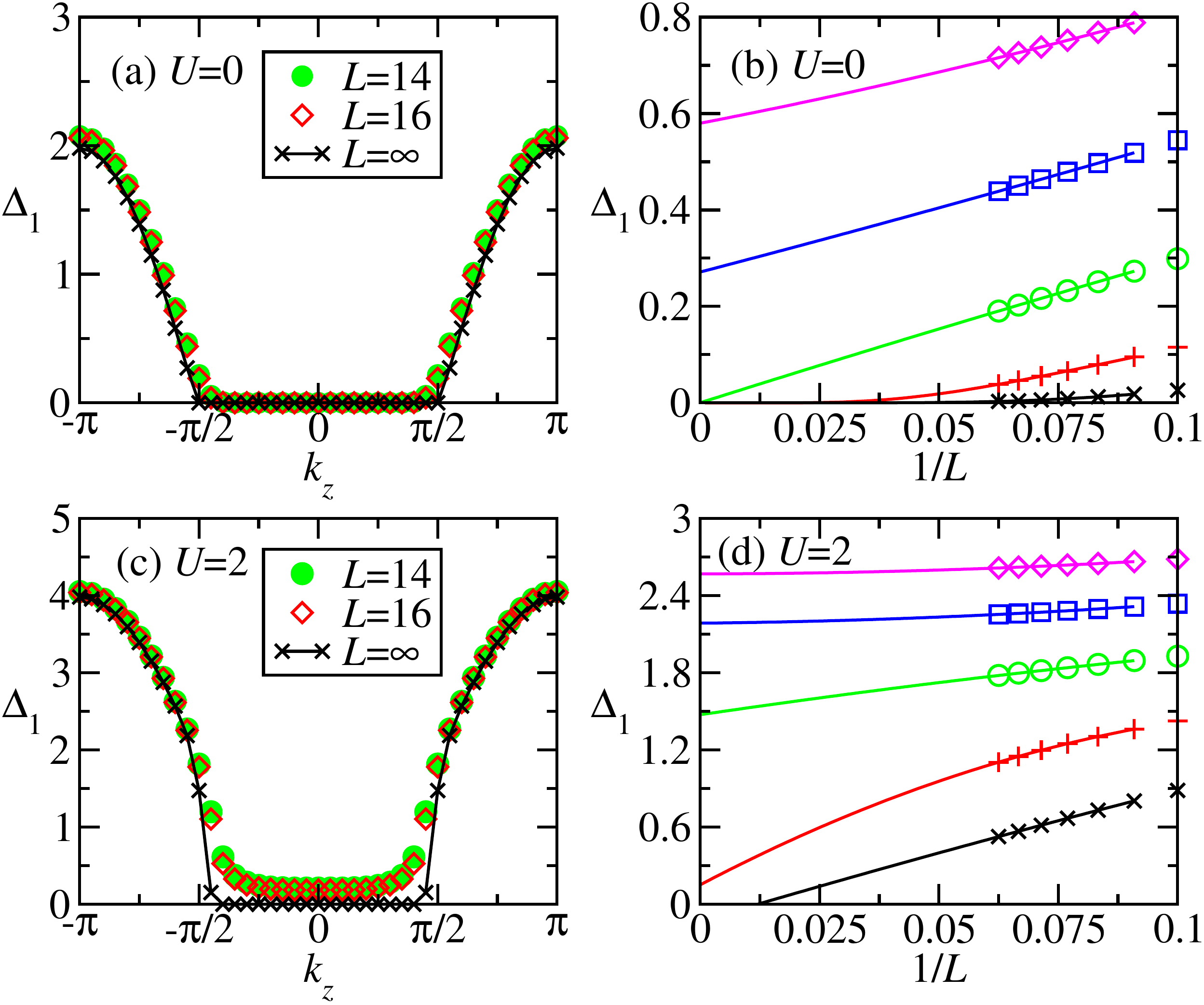}
\caption{Size dependence of the single-particle gap $\Delta_1=\epsilon_+ - \epsilon_-$ as functions of $k_z$ for (a) $U=0$ and (c) $U=2$. The extrapolation of $\Delta_1$ to the thermodynamic limit for $k_z=0.6\pi$, $0.55\pi$, $0.5\pi$, $0.45\pi$, and $0.4\pi$ from top to bottom are shown in (b) and (d), respectively. Here we take $M=2$ (thus $k_\mathrm{Weyl}=\pi/2$) and $k_y=0$.
We fit the data in (b) and (d) with a quadratic function $f(L)=a_0+a_1(1/L)+a_2(1/L)^2$, except those of $|k_z|<\pi/2$ in (b), in which the exponential function $f(L)=a\;\mathrm{exp}(-bL)$ is employed.
} \label{single_gap}%
\end{figure}

After turning on the interaction $U$, the conclusions are qualitatively unchanged. The calculated quasi-particle/quasi-hole energies for $k_z=3\pi/4>k_\mathrm{Weyl}$ agree well with their bulk values $\epsilon^\mathrm{bulk}_\pm$ and thus no edge state appears. Interestingly, when $k_z=\pi/4<k_\mathrm{Weyl}$, the edge modes around $k_y=0$ are not destroyed by interactions. Nevertheless, the correlation effect makes the slopes of their dispersions around $k_y=0$ being steeper. This is consistent with the enhanced bulk gap due to electronic correlations. We note that the calculated single-particle gap $\Delta_1$ defined by the minimal value of the energy difference $\epsilon_+-\epsilon_-$ can have a tiny nonzero value due to finite-size effects. As shown below (Fig.~\ref{single_gap}), this value approaches zero in the thermodynamic limit, and then the zero-energy edge modes indeed show up in the spectrum. Therefore, our findings clearly demonstrate the robustness of zero-energy single-particle edge modes against the long-ranged interaction $U$.

\begin{figure}[tp]
\includegraphics[width=0.48\textwidth]{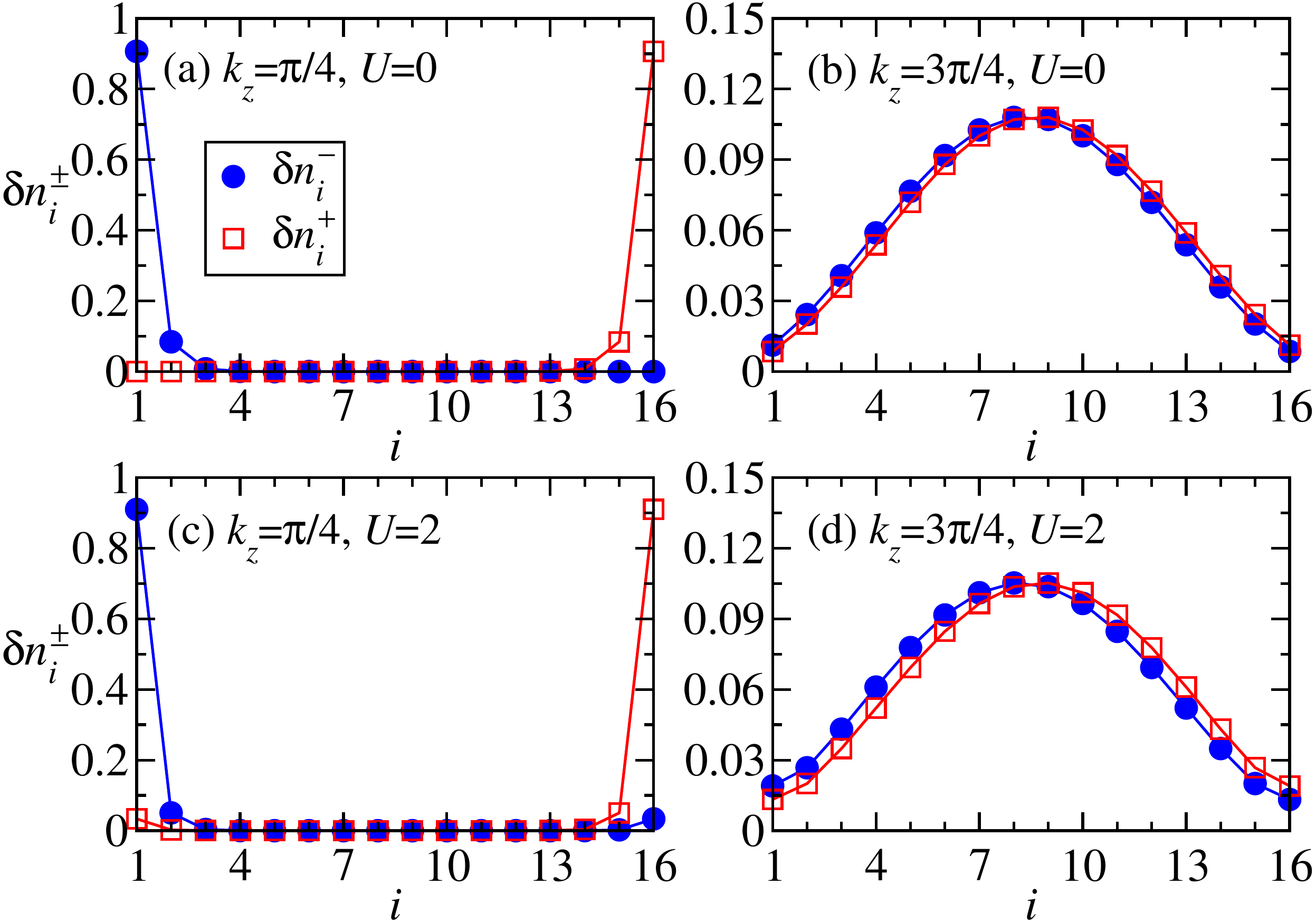}
\caption{Density profiles $\delta n_i^\pm$ at two different $U$'s of an extra particle or hole for the states with $k_y=0.05\pi$ for both the cases of $k_z=\pi/4$ and $3\pi/4$. Here $L=16$. The upper panels show the $U=0$ cases, while the lower ones gives those for $U=2$.}\label{den_profile}
\end{figure}

To further identify the in-gap states around $k_y=0$ for $k_z=\pi/4$ as edge modes, we need to show that the densities of these quasi-particle/quasi-hole states do distribute near the edges only. The density profile of one electron added is defined as %
$\delta n_i^+=\langle\tilde{n}_{i}\rangle_{N=L+1} -\langle\tilde{n}_{i}\rangle_{N=L}$ and that of an extra hole is $\delta n_i^-=\langle\tilde{n}_{i}\rangle_{N=L} -\langle\tilde{n}_{i}\rangle_{N=L-1}$. Here $\tilde{n}_{i}=\sum_\alpha\tilde{c}_{i\alpha}^\dagger\tilde{c}_{i\alpha}$ is the total electron number operator on site $i$ and $\langle\tilde{n}_{i}\rangle_{N}$ denotes its expectation value with respect to the ground state with $N$ electrons. The values of $\delta n_i^+$ and $\delta n_i^-$ describe how an added particle or hole distributes itself in the system.

Our findings of $\delta n_i^+$ and $\delta n_i^-$ at $U=2$ for the states with a typical value of $k_y$ around $k_y=0$ are shown in Fig.~\ref{den_profile} for both the cases of $k_z=\pi/4$ and $3\pi/4$. Again, the data for the $U=0$ case are displayed as well for comparison. When $k_z=3\pi/4>k_\mathrm{Weyl}$, no matter whether the interaction $U$ is present or not, the extra particle/hole always distributes itself in the center of the lattice system. It shows that these quasi-particle/quasi-hole states indeed describe the bulk states, in support of the energy perspective discussed above. For $k_z=\frac{\pi}{4}<k_\mathrm{Weyl}$, we find that the added particle/hole locates on one or the other edge even in the presence of the long-ranged interaction $U$. Moreover, the density profiles for the $U=2$ case modify little as compared to the noninteracting case. Therefore, the in-gap states for $k_z=\pi/4$ do behave as edge modes and they are stable against electronic correlations.

The Fermi arc is constituted by the collection or locus of the zero-energy edge states. In the noninteracting limit, this line segment ends at the projection of the bulk Weyl nodes onto the surface Brillouin zone. It has been shown in Fig.~\ref{1part_eng} that, for a specific $k_z=\pi/4$, the zero-energy edge modes, indicated by vanishing single-particle gap $\Delta_1$, persist upon the interaction effect. We now extend such discussions to other $k_z$'s and determine the whole Fermi arc for the present interacting model.

In Fig.~\ref{single_gap}, the single-particle gaps $\Delta_1$ for all $k_z$'s within the Brillouin zone are displayed for the case of $U=2$. For comparison, the results for the noninteracting ($U=0$) case are shown as well. As illustrated in Fig.~\ref{1part_eng}, the possible zero-energy modes always occur at $k_y=0$. We thus fix $k_y=0$ in our calculations of $\Delta_1$. Besides, $M=2$ is assumed such that $k_\mathrm{Weyl}=\pi/2$. The $U=0$ results can be derived analytically. When $|k_z|<\pi/2$, $\Delta_1=0$ because of the appearance of a pair of degenerate chiral edge mode at $k_y=0$ [cf. Fig.~\ref{1part_eng}(a)]. On the other hand, when $\pi/2<|k_z|\le\pi$, there exists no edge mode and the quasi-particle/quasi-hole energies agree well with their bulk values $\epsilon^\mathrm{bulk}_\pm$ such that  $\Delta_1=2\left[\epsilon^\mathrm{bulk}_+\right]_{k_y=0}$, which reduces to $\Delta_1=2|\cos{k_z}|$ for the present case of $M=2$. That is, the Fermi arc at $U=0$ extends from $k_z=-\pi/2$ to $\pi/2$. As seen from Fig.~\ref{single_gap}, our data agree with the analytic predictions, while visible deviation due to finite-size effects is observed around $k_z=\pm\pi/2$. Nevertheless, after extrapolating to the thermodynamic limit with $L\rightarrow\infty$, the agreement becomes excellent.

When the long-ranged interaction $U$ is turned on, for $\pi/2<|k_z|\le\pi$, the calculated $\Delta_1$ again agrees well with the analytic prediction $\Delta_1=2\left[\epsilon^\mathrm{bulk}_+\right]_{k_y=0}=2|\cos{k_z}|+U$. However, for $|k_z|\leq\pi/2$, enhanced finite-size effects are observed in our calculations~\cite{note_gap}. While $\Delta_1$ is found to be always nonzero up to our maximum size $L=16$, its extrapolated values for $|k_z|\le0.4\pi$ does reduce to zero. We note that the edge states are expected to be more and more delocalized when $k_z$ approaches the Weyl nodes at $k_z=\pm\pi/2$. Therefore, severe finite-size effects exist around the Weyl nodes and calculations for much larger sizes are necessary to reduce such effects. Nevertheless, our findings strongly suggest that the Fermi arcs for nonzero $U$ have the same extent as those in the noninteracting case. We thus conclude that, except the renormalized dispersions, the single-particle Fermi arcs receive no modification by the long-ranged interaction.

It appears that our conclusion disagrees with that obtained in Ref.~\cite{Meng-Budich2019}, where a gap opening in the single-particle Fermi-arc states is observed for a slightly different model. Actually, after a necessary correction to their model Hamiltonian, gapless single-particle Fermi arcs will be recovered. That is, the same conclusion can be reached even when other long-ranged interactions different from the present model are taken.
Instead of Eq.~\eqref{eq:1D_HU}, the authors in Ref.~\cite{Meng-Budich2019} consider the following interaction term to simplify their calculations,
\begin{equation}\label{eq:Meng-Budich_HU}
\widetilde{\mathcal{H}}_U (\mathbf{k}_\bot) =
\frac{U(k_z)}{2} \left[ \sum_{i=1}^L ( \tilde{n}_{i\uparrow} + \tilde{n}_{i\downarrow} -1 ) \right]^2 \; .
\end{equation}
Because this long-ranged density-density interaction depends only on the total particle number $\tilde{N}=\sum_{i=1}^L(\tilde{n}_{i\uparrow}+\tilde{n}_{i\downarrow})$, it is clear that both the quasi-particle and the quasi-hole energies, $\epsilon_+$ and $\epsilon_-$, will be pushed away from the Fermi level by $U(k_z)/2$ for those $k_z$'s with nonzero $U(k_z)$'s. This gives a gap opening of $U(k_z)$ for the noninteracting Fermi-arc states.
However, the form in Eq.~\eqref{eq:Meng-Budich_HU} is improper physically. For many-body states with a fixed particle density $\tilde{N}/L=1\pm\delta\rho$, where $\delta\rho$ denotes the density deviation from half-filling, this interaction term gives a contribution in energy density, $U(k_z)(\delta\rho)^2 L/2$, which becomes infinite in the thermodynamic limit, $L\rightarrow\infty$.
In order to keep the energy density being an intensive quantity, a factor $1/L$ needs to be added in Eq.~\eqref{eq:Meng-Budich_HU}, i.e., $\widetilde{\mathcal{H}}_U\rightarrow\widetilde{\mathcal{H}}_U/L$. The single-particle gap thus becomes $U(k_z)/L$, which vanishes in the thermodynamic limit. Therefore, the corrected results are actually in agreement with ours.

\subsection{Collective Fermi arcs of particle-hole nature}

While the Weyl nodes are gapped out by the long-ranged interactions, the Hall conductance (and thus the Chern number) for each 2D momentum sector between the nodes remains its quantized value~\cite{Morimoto-Nagaosa2016}. According to the theory of quantum Hall edge states~\cite{Wen1991,Renn1995,Yoshioka}, there must exist chiral edge modes on the boundaries of these 2D sectors. Notice that, instead of quasiparticle excitations, those edge modes are the capillary waves, whose dynamics is described by a 1D chiral boson theory. That is, the Fermi arc states implied by nonzero Chern number are constituted by the collective edge excitations, rather than the single-particle ones. Since the topological Chern number of the bulk system is not affected by the long-ranged interactions, the collective Fermi arcs of particle-hole nature are expected to maintain their noninteracting structures.

To verify this conclusion, we evaluate the particle-hole excitation energies $\Delta\mathcal{E}_\mathrm{ph}=\mathcal{E}_1-\mathcal{E}_0$ for the cases of nonzero $U$'s. Here $\mathcal{E}_0$ ($\mathcal{E}_1$) denotes the energy of the ground state (first-excited state) for the 1D model in Eq.~\eqref{eq:1D_H} at half-filling. Because the particle number is fixed, the excited states should represent the particle-hole excitations. The appearance of the edge modes can be recognized once $\Delta\mathcal{E}_\mathrm{ph}$ deviates the corresponding bulk value. From the discussions at the end of Sec.~\ref{bulk}, the lowest excitation energy of the bulk particle-hole excitation at zero momentum transfer for a given set of $k_y$ and $k_z$ can be obtained by
\begin{align}
\Delta E_\mathrm{ph}^\mathrm{bulk} &=\min_{k_x}\{2h(\mathbf{k})\} \nonumber \\ &=2\sqrt{\sin^2{k_y}+(1-\cos{k_y}-\cos{k_z})^2} \nonumber
\end{align}
for the present case of $M=2$. This expression is valid no matter whether $U$ is zero or not. We note that, different from the cases of $\epsilon^\mathrm{bulk}_\pm$, the bulk value of the particle-hole excitation energy $\Delta E_\mathrm{ph}^\mathrm{bulk}$ can approach zero at the Weyl nodes with $k_y=0$ and $k_z=\pm\pi/2$, even when $U$ is nonzero.

\begin{figure}[tp]
\includegraphics[width=0.48\textwidth]{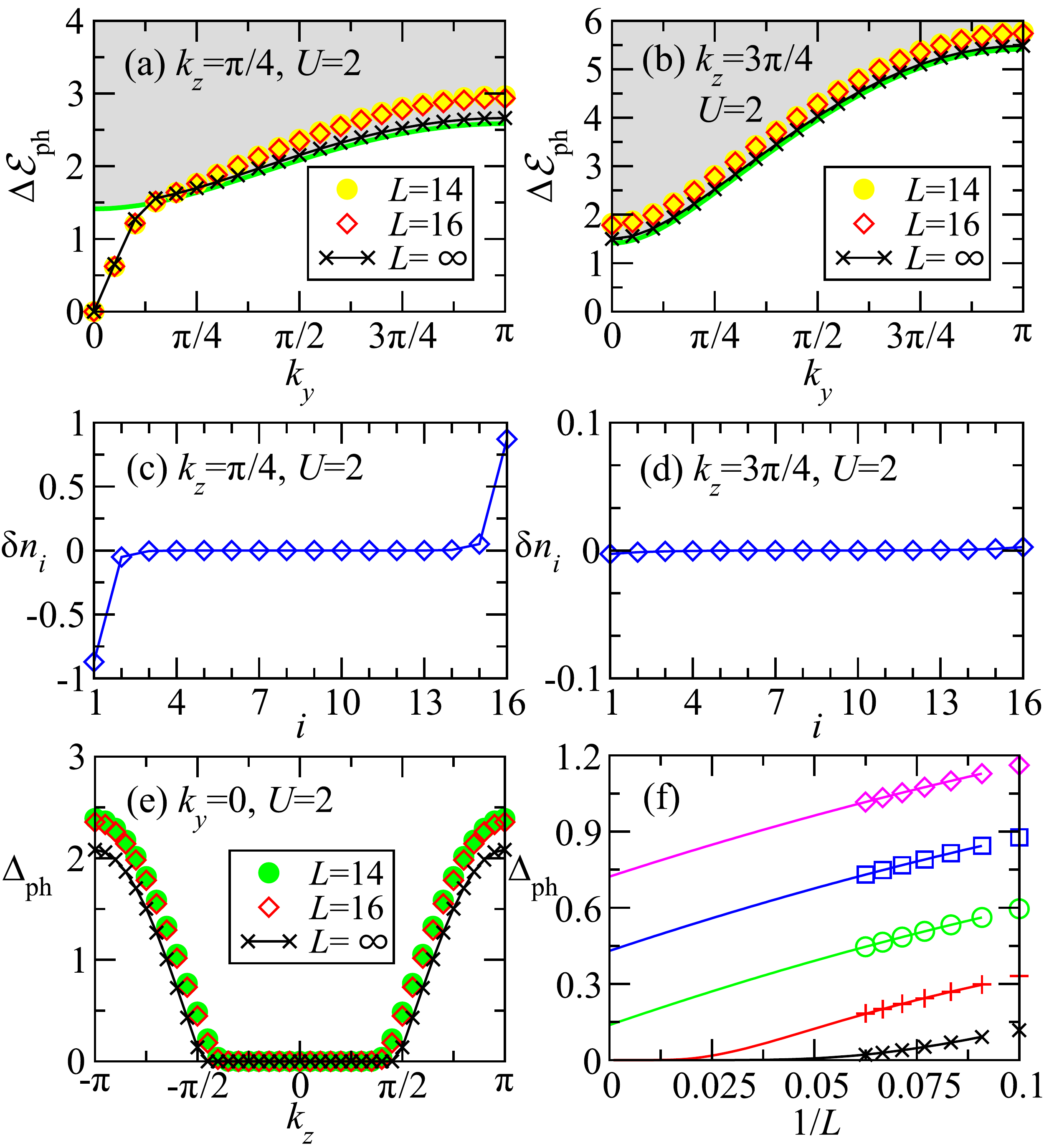}
\caption{Size dependence of particle-hole excitation energies $\Delta\mathcal{E}_\mathrm{ph}$ at $U=2$ as functions of $k_y$ for (a) $k_z=\pi/4$($<k_\mathrm{Weyl}$) and (b) $3\pi/4$($>k_\mathrm{Weyl}$). Here we take $M=2$ such that $k_\mathrm{Weyl}=\pi/2$. The green lines denote the corresponding bulk values $\Delta E_\mathrm{ph}^\mathrm{bulk}$.
Panels (c) and (d) show the density profiles $\delta n_i$ of the particle-hole excitations at $U=2$ with $k_y=0.05\pi$ for the cases of $k_z=\pi/4$ and $3\pi/4$, respectively. Here the data for $L=16$ are presented.
Panel (e) displays the size dependence of the particle-hole gap $\Delta_\mathrm{ph}$ as functions of $k_z$ for $U=2$. The extrapolation of $\Delta_\mathrm{ph}$ to the thermodynamic limit for $k_z=0.6\pi$, $0.55\pi$, $0.5\pi$, $0.45\pi$, and $0.4\pi$ from top to bottom is shown in (f).
The data in (f) are fitted with a quadratic function $f(L)=a_0+a_1(1/L)+a_2(1/L)^2$, except those of $|k_z|<\pi/2$, in which the exponential function $f(L)=a\;\mathrm{exp}(-bL)$ is employed.
}\label{part_hole}
\end{figure}

Our results of $\Delta\mathcal{E}_\mathrm{ph}$ as functions of $k_y$ for $k_z=\pi/4$ and $3\pi/4$ are shown in Figs.~\ref{part_hole}~(a) and (b) for the $U=2$ case. Because of the symmetry in the energy spectrum mentioned in the last subsection, only the data for $0\leq k_y<\pi$ are plotted. As illustrated in Fig.~\ref{part_hole}~(a), we find that the edge modes do appear around $k_y=0$ for $k_z<k_\mathrm{Weyl}$ even in the presence of long-ranged interactions. Notably, the calculated energies of bulk excitations for either $k_z<k_\mathrm{Weyl}$ or $k_z>k_\mathrm{Weyl}$ experience considerable finite-size effects. Nevertheless, their values agree well with the analytic predictions after extrapolating to the thermodynamic limit.

To show the in-gap excited states around $k_y=0$ for $k_z=\pi/4$ being the edge modes, their density profiles $\delta n_i=\langle\tilde{n}_{i}\rangle_{1} -\langle\tilde{n}_{i}\rangle_{0}$ are calculated as well. Here $\langle\cdots\rangle_{0}$ and $\langle\cdots\rangle_{1}$ denote the expectation values with respect to the ground state and the first-excited state at half-filling, respectively. Our result at $U=2$ for the state with $k_z=\pi/4$ and a typical value of $k_y$ around $k_y=0$ is displayed in Fig.~\ref{part_hole}~(c). For comparison, the counterpart for the case of $k_z=3\pi/4$ is plotted in Fig.~\ref{part_hole}~(d). As illustrated in the figure, the in-gap states around $k_y=0$ for $k_z=\pi/4$ do behave as edge states, in agreement with the energy perspective. Besides, the disturbed density shows a nonlocal charge transfer from one end to the other. That is, this particle-hole excitation implies the quantized charge pumping via the twisted boundary condition in the $y$ direction with a twist angle $\phi=2\pi$. Such a charge pumping process is expected, since the 2D sector at $k_z=\pi/4$ behaves as a quantum Hall system with quantized Hall conductance.

Finally, we determine the extent of the collective Fermi arcs constituted by the collection of the zero-energy edge states of particle-hole excitations. Similar to the discussions in Fig.~\ref{single_gap}, this can be achieved by evaluating the particle-hole gap $\Delta_\mathrm{ph}$ defined by the minimal value of $\Delta\mathcal{E}_\mathrm{ph}$. Therefore, the locus of the states with $\Delta_\mathrm{ph}=0$ in the surface Brillouin zone gives the collective Fermi arc. As observed in Figs.~\ref{part_hole}~(a) and (b), the states with the smallest $\Delta\mathcal{E}_\mathrm{ph}$ always occur at $k_y=0$. We thus fix $k_y=0$ in our calculations of $\Delta_\mathrm{ph}$~\cite{note2}. Here $k_\mathrm{Weyl}=\pi/2$ because $M=2$ is assumed.
Our findings of $\Delta_\mathrm{ph}$ as functions of $k_z$ for the case of $U=2$ are displayed in Fig.~\ref{part_hole}~(e). Because $\Delta_\mathrm{ph}=\Delta_1$ for $U=0$, the results of $\Delta_\mathrm{ph}$ for the noninteracting case can directly refer to Fig.~\ref{single_gap}~(a). We find that the zero-energy edge states of particle-hole excitations appear when $k_z$ lies between two Weyl nodes even in the presence of long-ranged interactions. As compared to the single-particle counterparts in Fig.~\ref{single_gap}~(c), the particle-hole gaps within $|k_z|<\pi/2$ are found to receive less finite-size effects~\cite{note_gap}. As mentioned before, the occurrence of such collective zero-energy edge states is actually guaranteed by the nonzero Hall conductances. Our findings provide a numerical support on this bulk-boundary correspondence in terms of collective excitations.

\section{conclusions and discussions}

In summary, by means of numerical exact diagonalizations, we provide clear evidences on the existence of the Fermi-arc surface states even when the bulk Weyl nodes are gapped out by the long-ranged interactions. That is, the relationship between the existence of the gapless Weyl nodes and the surface Fermi arcs needs not hold in the correlated systems, while it is always true in the noninteracting cases.
The possibility of the appearance of Fermi arcs in the absence of gapless nodes exhibit the subtlety in identifying the topological phases of interacting WSMs. For example, the conclusions drawn merely from the nodal structure of the single-particle spectrum should be considered with caution.

We note that the Fermi-arc surface states can be defined both in the $\Delta N=\pm 1$ and the $\Delta N=0$ subspaces of the Hilbert space, where $\Delta N$ denotes the change in total particle number. The former is of quasiparticle nature, while the latter describes the collective particle-hole excitation. For the model under consideration, both kinds of the Fermi-arc surface states are found to preserve their noninteracting structures even in the presence of the long-ranged interactions.

Due to their topological origin, the robustness of the collective Fermi-arc surface states against electronic correlations is actually expected. It is shown that the bulk Hall conductance (and the Chern number) of each 2D momentum sector still keep its noninteracting value even though the long-ranged interactions produce single-particle gaps at the Weyl nodes~\cite{Morimoto-Nagaosa2016}. According to the bulk-boundary correspondence given by the theory of quantum Hall edge states~\cite{Wen1991,Renn1995,Yoshioka}, the collective chiral edge states and then the constituted Fermi arcs are guaranteed to appear on the boundaries.

On the other hand, the existence of the single-particle Fermi-arc states established by our numerical calculations may be somehow surprising. The appearance of such single-particle states cannot be directly inferred from the results of Hall conductance. This statement is based on the fact that the Hall conductance describes the current-current correlator and thus contains only the information of the particle-hole subspace, rather than the quasiparticle one. While the presence of single-particle Fermi arcs was suggested by solving the effective single-particle Hamiltonian $H_t$~\cite{Morimoto-Nagaosa2016}, such a conclusion can be misleading because the eigenstates of $H_t$ have no direct physical meaning~\cite{Gurarie2011,Essin-Gurarie2011}. Therefore, to prove their existence, one has to diagonalize the many-body Hamiltonian within the single-particle/hole subspaces, as done in the present work.

Our results are achieved for the WSMs with long-ranged interactions. Some comments on their validity in the cases of generic interactions are made as follows.
While the single-particle Fermi arcs are shown to be robust against the long-ranged interactions irrespective of the single-particle gap opening, their stability may not hold for generic interactions. Recall that the subspaces of different momenta are decoupled in the present model with momentum-local interactions. Such decoupling will not happen when the momentum-local condition is relaxed. Let's consider a more realistic model such that the interactions in the two directions perpendicular to the $z$ axis, along which a pair of Weyl nodes is separated, become short-ranged. Therefore, states with different $(k_x,\,k_y)$ get coupled. For such a model, the region between nodes can still be viewed as a stack of 2D interacting Chern insulators, if the Hall conductances of these 2D sectors remain quantized. After imposing the open boundary conditions, because of the couplings among states with different momenta along the boundaries, the effective model on the edges of these 2D systems will be described by the Luttinger-liquid theory~\cite{Wen1991,Renn1995,Yoshioka}. This theory shows that the quasiparticle edge modes become gapped and the gapless edge modes must consist of the particle-hole collective excitations. As a consequence, we conclude that, in the generic cases, the single-particle Fermi arcs can disappear, even though the collective ones remain topologically protected by nonzero Chern numbers.
In addition, we stress that the stability of the collective Fermi arcs relies only on the topology in the particle-hole subspace and is irrelevant to the fact whether the single-particle gaps at the nodes are opened or not. That is, it is possible to have interacting WSMs hosting gapped nodes and with collective Fermi arcs on their surfaces.

Due to distinct characters in the single-particle and the collective Fermi arcs, one needs to measure different physical quantities to determine their existence. For example, the angle-resolved photoemission spectroscopy, which measures the single-particle spectrum functions, can be employed to observe only the single-particle Fermi arcs. To detect directly the collective Fermi arcs consisting of particle-hole excitations, more delicate arrangements are necessary. Recently, experimental techniques to reveal the momentum and energy dependent two-particle correlations are proposed~\cite{Su_Zhang20}. Due to the advances in experiments, it may be possible to observe collective Fermi arcs in the near future and then to verify our conclusions.

\begin{acknowledgments}
We are grateful to Tobias Meng for useful discussions.
Y.C.T. and M.F.Y. acknowledge support from the Ministry of Science and Technology of Taiwan under Grant No. MOST 108-2112-M-029-005 and  108-2811-M-005-522, respectively.
\end{acknowledgments}

%


\begin{thebibliography}{34}
\bibitem [{not({\natexlab{b}})}]{note_gap}%
  \BibitemOpen
  \href@noop {} {} {\natexlab{b}}\bibinfo {note} {%
  In the noninteracting case, the finite-size gap of the Fermi-arc states is expected to vanish exponentially with the system size $L$. This is because the gap opening comes from the mixing of the edge states at opposite ends and these edge modes decay into the bulk exponentially. Our data of the single-particle gap $\Delta_1$ at $U=0$ for $|k_z|<\pi/2$ agree well with this conclusion. While this size dependence may apply as well to the systems under short-ranged interactions, it could not be true for the present model with a long-ranged interaction $U$. Indeed, we find that $\Delta_1$ becomes to vanish linearly/polynomially with $1/L$ (up to our largest size $L=16$) for $|k_z|\leq0.4\pi$, as indicated in Fig.~\ref{single_gap}~(d) for $k_z=0.4\pi$. For those $k_z$'s near $k_z=\pm0.5\pi$ (say, $k_z=0.45\pi$), conclusive results can be achieved only when numerical calculations for much larger sizes become available.
  %
  Interestingly, it is found that the finite-size gap $\Delta_{\mathrm{ph}}$ of the particle-hole edge excitations for $|k_z|<\pi/2$ always vanish exponentially with $L$, no matter whether $U$ is zero or not. We note that this $U$-independence appears as well in the bulk particle-hole exciation energies $\Delta E_\mathrm{ph}$, as mentioned at the end of Sec.~\ref{bulk}.
  }\BibitemShut {Stop}%
\bibitem [{not({\natexlab{c}})}]{note2}%
  \BibitemOpen
  \href@noop {} {} {\natexlab{c}}\bibinfo {note} {In this case of $k_y=0$,
  the 1D Hamiltonian $\mathcal{H}(\mathbf{k}_\perp )$ in Eq.~\eqref{eq:1D_H} is invariant
  under the parity transformation $\mathcal{P}$ discussed in the subsection~\ref{1p_arc}.
  The eigenvalues of the parity (i.e., $\mathcal{P}=\pm1$ because
  $\mathcal{P}^2=1$) thus become good quantum numbers and the diagonalization
  can be performed for each parity sector. $\Delta_{\mathrm{ph}}$ is obtained
  by taking the energy difference between these two parity sectors.}\BibitemShut {Stop}%
\end{thebibliography}
\end{document}